\documentclass[11pt]{article}

\usepackage{graphicx,graphics,color,amsmath}
\newcommand{\rem}[1]{}

\newsavebox{\astrutbox}
\sbox{\astrutbox}{\rule[-5pt]{0pt}{20pt}}

\newcommand\etal{\mbox{\textit{et al.~}}}

\DeclareMathAlphabet{\mathbi}{OML}{cmm}{b}{it}

\newcommand{\bx}{\mathbi{x}}

\newcommand{\bel}{\begin{equation}\label}
\newcommand{\ee}{\end{equation}}
\newcommand{\beq}{\begin{eqnarray}\label} 
\newcommand{\eeq}{\end{eqnarray}} 
\newcommand{\bc}{\begin{center}} 
\newcommand{\ec}{\end{center}} 
\newcommand{\ben}{\begin{enumerate}}
\newcommand{\een}{\end{enumerate}}
\newcommand{\bit}{\begin{itemize}}
\newcommand{\eit}{\end{itemize}}
\newcommand{\I}{\int_{\mathcal{V}}}
\newcommand{\bom}{\mbox{\boldmath$\omega$}}
\newcommand{\tom}{\tilde{\Omega}}
\newcommand{\bdf}{\mathbi{f}}

\newcommand{\bu}{\mbox{\boldmath$u$}}

\newcommand{\shalf}{{\ensuremath{\scriptstyle\frac{1}{2}}}}
\makeatletter
\@addtoreset{equation}{section}

\makeatother

\begin{document}
\sf\bc
\textbf{\color{blue}High-low frequency slaving and regularity issues in the $3D$ Navier-Stokes equations}
\par\vspace{3mm}
\textbf{J. D. Gibbon\footnote{\sf j.d.gibbon@ic.ac.uk and www2.imperial.ac.uk/$\sim$jdg} }
\par\vspace{2mm}
Department of Mathematics, Imperial College London, SW7 2AZ, UK
\ec

\begin{abstract}
The old idea that an infinite dimensional dynamical system may have its high modes or frequencies slaved to low modes 
or frequencies is re-visited in the context of the $3D$ Navier-Stokes equations. A set of dimensionless frequencies 
$\{\tom_{m}(t)\}$ are used which are based on $L^{2m}$-norms of the vorticity. To avoid using derivatives a closure 
is assumed that suggests that the $\tom_{m}$ ($m>1$) are slaved to $\tom_{1}$ (the global enstrophy) in the form $\tom_{m} 
= \tom_{1}\mathcal{F}_{m}(\tom_{1})$.  This is shaped by the constraint of two H\"older inequalities and a time average from 
which emerges a form for $\mathcal{F}_{m}$ which has been observed in previous numerical Navier-Stokes and MHD simulations. 
When written as a phase plane in a scaled form, this relation is parametrized by a set of functions $1 \leq \lambda_{m}(\tau) \leq 4$, 
where curves of constant $\lambda_{m}$ form the boundaries between tongue-shaped regions. In regions where 
$2.5 \leq \lambda_{m} \leq 4$ and $1 \leq \lambda_{m} \leq 2$ the Navier-Stokes equations are shown to be regular\,: 
numerical simulations appear to lie in the latter region. Only in the central region $2 < \lambda_{m} < 2.5$ has no proof 
of regularity been found.
\end{abstract}
\bc
\textit{\color{magenta}Dedicated to the memory of David Broomhead (1950 -- 2014)}
\ec

\section{\textsf{\color{blue}Introduction}}\label{intro}

\subsection{\textsf{\color{blue}Historical background}}\label{hb}

A generation ago a recurrent theme in studies in infinite dimensional dynamical systems was the idea that a 
small subset of low modes or coherent states might conceivably control the dynamics by slaving the higher 
modes to this subset. Having originally emerged from earlier work on centre manifolds in studies in ordinary 
differential equations (Broomhead, Indik, Newell and Rand (1991), Guckenheimer and Holmes (1997), Holmes, 
Lumley and Berkooz (1996)), such ideas obviously have a lasting appeal, particularly for those who work in 
turbulent flows where the number of degrees of freedom are so large that resolved computations at realistic 
Reynolds numbers are hard to achieve\,: see Moin and Mahesh (1998), Donzis, Young and Sreenivasan (2008), 
Pandit, Perlekar and Ray (2009), Ishihara, Gotoh and Kaneda (2009), Kerr (2012,\,2013) and Schumacher, Scheelb, 
Krasnov, Donzis, Yakhot and Sreenivasan (2014). For partial differential equations there also emerged a 
parallel and closely related body of work on global attractors and inertial manifolds which aimed to prove the 
finite dimensionality of the system in question in some specified sense\,: Foias, Sell and Temam (1988), Titi 
(1990), Foias and Titi (1991), Robinson (1996), Foias, Manley, Rosa and Temam (2001). Some success was 
achieved for the one-dimensional  the Kuramoto-Sivashinsky equation where an inertial manifold was 
proved to exist (Foias, Jolly, Kevrekidis, Sell and Titi 1988). Further success was also achieved for the $2D$ 
incompressible Navier-Stokes equations when a global attractor was shown to exist with a sharp estimate 
for its dimension (Constantin, Foias and Temam 1988), with further estimates on the number of determining 
modes and nodes\,: see Foias and Prodi (1967), Foias and Temam (1984), Jones and Titi (1993), Olson and Titi 
(2003) and Farhat, Jolly and Titi (2014).  
\par\smallskip
The aim of this paper is to revisit the low/high mode idea in a new way in the context of the open question of 
the global regularity of the $3D$ Navier-Stokes equations
\begin{equation}\label{nse1}
\partial_{t}\bu + \bu\cdot\nabla\bu = \nu \Delta\bu - \nabla p + \bdf(\bx)\,,\qquad\qquad \mbox{div}\,\bu = \mbox{div}\,\bdf =0\,,
\end{equation} 
with periodic boundary conditions on a cube of volume $\mathcal{V} = [0,\,L]_{per}^{3}$. The body force $\bdf(\bx)$ is taken 
to be $L^{2}$-bounded and centred around a forcing length scale $\ell$ in the manner described by Doering and Foias (2002)\,: 
in this paper $\ell$ is taken as $\ell = L$ for convenience. The two dimensionless numbers corresponding to the forcing and the 
system response are respectively given by the Grashof and Reynolds numbers
\begin{equation}\label{GRdef}
Gr = \frac{L^{3/2}\|\bdf\|_{2}}{\nu^2}\,\qquad\qquad Re = \frac{L U_{0}}{\nu}\,
\end{equation}
where $U_{0}^{2} = L^{-3}\left<\|\bu\|_{2}^{2}\right>_{T}$ with $\|\cdot\|_{2}$ representing the $L^2$-norm and $\left<\cdot\right>_{T}$ 
a time average up to time $T > 0$. 

\subsection{\textsf{\color{blue}Some recent analytical and numerical scaling results}}\label{scaling}

Given the open nature of the question of global regularity of solutions of the $3D$ Navier-Stokes equations, analyses have 
conventionally been based on an assumption of some type. The main one has been that the velocity field is assumed to 
remain bounded in some function space, the best being the $u \in L^{3}(\mathcal{V} )$ result of Escauriaza, Seregin and 
Sver\'ak (2003). An alternative computational approach has been to discuss low modes in terms of coherent states by 
projecting onto special selections of Fourier-Galerkin modes (Broomhead, Indik, Newell and Rand (1991), Holmes, Lumley 
and Berkooz (1996)). 
\par\smallskip
Here we break with both of these traditions and instead introduce a set of time dependent frequencies or inverse time 
scales (or `modes')
\bel{Omegaset}
\{\Omega_{1}(t),\,\Omega_{2}(t),\,\ldots\,\Omega_{m}(t)\}\,,
\ee
based on $L^{2m}$-norms of the three-dimensional vorticity field $\bom(\bx,\,t)$ which obeys 
\bel{}
\left(\partial_{t} + \bu \cdot\nabla\right)\bom = \nu \Delta\bom + \bom\cdot\nabla\bu + \mbox{curl}\,{\bdf}\,.
\ee
The $\Omega_{m}$ in (\ref{Omegaset}) are defined such that each has the dimension of a frequency
\bel{f1}
\Omega_{m}(t) = \left(L^{-3}\I |\bom|^{2m}\,dV\right)^{1/2m}\,.
\ee
$\Omega_{1}(t)$ is the global enstrophy and the $\Omega_{m}(t)$ are higher moments. The set (\ref{Omegaset}) would be 
widely spread if the vector field $\bom(\bx,\,t)$ is strongly intermittent, whereas they would be squeezed closely together 
if $\bom$ is mild in behaviour. Multiplication by the inverse of the constant frequency $\varpi_{0}= \nu L^{-2}$ produces a 
non-dimensional set $\tom_{m} = \varpi_{0}^{-1}\Omega_{m}$. 
\par\smallskip
In earlier work, a scaled set of the $\tom_{m}$ 
\bel{Dmdef1}
D_{m} = \tom_{m}^{\,\alpha_{m}}\qquad\mbox{with}\qquad\alpha_{m} = \frac{2m}{4m-3}\,,
\ee
was shown to have bounded time averages for $1 \leq m \leq \infty$ (Gibbon 2011)
\bel{timav1}
\left<D_{m}\right>_{T} \leq c\,Re^3 + O\left(T^{-1}\right)\,.
\ee
Given that $\alpha_{1}=2$, the first of these, $\left<D_{1}\right>_{T}\leq c\,Re^3$, is just the well-known result that the time 
average of the global enstrophy is a bounded quantity. The origin of the $\alpha_{m}$-scaling in (\ref{Dmdef1}) comes from 
symmetry considerations. 
\par\smallskip
It was observed in Donzis \etal (2013) that time plots of the $D_{m}(t)$ from several different simulations were ordered on a 
descending scale. In a further paper (Gibbon \etal 2014), it was observed that plots of the maxima in time of $\ln D_{m} / 
\ln D_{1}$ led to the relation
\bel{f7}
D_{m} \leq D_{1}^{A_{m,\lambda}}\qquad\mbox{with}\qquad A_{m,\lambda} = \frac{(m-1)\lambda + 1}{4m-3}\,,
\ee
where the accuracy of the fit lay within $5\%$.  The corresponding fixed values of the fitting parameter $\lambda$ lay in the 
range $1.15 \leq \lambda \leq 1.5$. These numerical simulations were\,: (i) a $1024^2\times 2048$ decaying calculation with 
anti-parallel initial conditions at about $Re_{\lambda} \sim 400$ based on work reported in Kerr (2012,\,2013)\,;  (ii) a forced 
and a decaying $(512)^3$ pair of simulations at about $Re_{\lambda} \sim 250$ -- see Gibbon \etal (2014)\,; (iii) data from a 
large-scale statistically steady simulation $(4096)^3$ simulation on $10^5$ processors at $Re_{\lambda} \approx 1000$, 
reported in Donzis \etal (2008,\,2010) and Yeung, Donzis and Sreenivasan (2012).  In addition to these, (\ref{f7}) has also been 
seen in a set of $3D$-MHD simulations in similar circumstances to those performed for the Navier-Stokes equations\,: see  
Gibbon \etal (2015).

\section{\textsf{\color{blue}A time-dependent closure avoiding derivatives of $\bom$}}

The inequality in (\ref{f7}) is a numerical observation. To put this, or something close to it, on a rigorous footing for the $3D$ 
Navier-Stokes equations, the first step is to assume that a strong solution\footnote{\sf We are using the standard contradiction 
method in PDE-analysis where it is assumed that there exists a maximal interval time $[0,\,T^{*})$ on which solutions of the 
Navier-Stokes equations exist and are unique\,: the strategy is to then attempt to prove a contradiction in the limit $t\to T^{*}$.} 
exists on a maximal time interval $[0,\,T^{*})$ on which it is assumed that $\tom_{1}$ is bounded -- hence all the other 
$\tom_{m}$ are also bounded. Then, instead of taking the standard route of a Sobolev inequality, which involves derivatives 
of $\bom$, 
the next step is to postulate a relation between $\tom_{m}$ and lower $\tom_{n}$ which could be thought of as a high/low 
frequency closure\footnote{\sf Strictly speaking, (\ref{f3}) is not a closure in the conventional sense used in turbulence modelling. 
Rather, it is an expression of how $\tom_{m}$  can be related to lower $\tom_{n}$,  under the constraints (\ref{Hin1}) and 
(\ref{Hin2}), without resorting to derivatives of $\bom$. Nevertheless, we will continue to use the word `closure' for convenience.}
\bel{f2}
\tom_{m} = F_{m}(\tom_{1},\, \tom_{2},\, \ldots\,,\, \tom_{n})\,,\qquad\qquad 1 \leq n < m\,.
\ee
This closure, which is designed to avoid the introduction of derivatives in $\bom$, must be constrained and shaped by the fact 
that $\tom_{m}$ must not only satisfy Holder's inequality
\bel{Hin1}
\tom_{1} \leq \ldots \leq \tom_{m} \leq \tom_{m+1}\,
\ee 
but it must also satisfy a triangular version H\"older's inequality  for $m > 1$ (see Appendix A)
\bel{Hin2}
\left(\frac{\tom_{m}}{\tom_{1}}\right)^{m^{2}} \leq \left(\frac{\tom_{m+1}}{\tom_{1}}\right)^{m^2 - 1}\,.
\ee
It is well known that the existence and uniqueness of solutions depends entirely on the bounded of the $H_{1}$-norm of the 
velocity field (Leray 1934), which is proportional to $\tom_{1}$. Thus we simplify the dependency of $\tom_{m}$ to the first 
frequency $\tom_{1}$ such that our `closure' in (\ref{f2}) is reduced to
\bel{f3}
\tom_{m} = \tom_{1}\mathcal{F}_{m}(\tom_{1})\,.
\ee
The inequalities (\ref{Hin1})  and (\ref{Hin2}) demand that $\mathcal{F}_{m}$ must satisfy both 
\bel{f4}
\mathcal{F}_{m+1} \geq \mathcal{F}_{m}^{\,m^{2}/(m^{2}-1)}\,,\qquad\qquad\mathcal{F}_{m} \geq 1\,.
\ee
The first inequality in (\ref{f4}) can be further simplified by the substitution
\bel{f5a}
\mathcal{F}_{m} = \left[h_{m}\left(\tom_{1},\tau\right) \right]^{\frac{m-1}{m}}
\ee
for a sequence of smooth, arbitrary functions $h_{m}\left(\tom_{1},\tau\right)$ which must then form a monotonically 
increasing sequence
\bel{f5b}
1 \leq h_{m}\left(\tom_{1},\tau\right) \leq  h_{m+1}\left(\tom_{1},\tau\right)\,.
\ee
In (\ref{f5a}) and (\ref{f5b}), $\tau$ is a dimensionless time $\tau = \varpi_{0}\,t$. Finally, (\ref{f3}) is reduced 
to
\bel{f5c}
\tom_{m} = \tom_{1}\left[h_{m}\left(\tom_{1},\tau\right)\right]^{\frac{m-1}{m}}\,.
\ee
The result depends only on the finiteness of the domain and the inverse box frequency $\varpi_{0} = \nu L^{-2}$ to create 
the dimensionless time $\tau$. The relation (\ref{f5c}) is no more than an expression of the potentially arbitrarily large 
distance between $\tom_{m}$ and $\tom_{1}$ in the form of an equality. In fact, (\ref{f5c}) contains no $3D$ Navier-Stokes 
information but once it is considered in this context, the finiteness of the time averages in (\ref{timav1}) must be enforced. 
An application of a H\"older inequality to (\ref{f5c}) shows that the $h_{m}$ must therefore be constrained by
\bel{hmcon1}
\left<h_{m}^{2/3}\right>_{T} < \infty\,.
\ee
One choice of $h_{m}$ consistent with this is that it cannot be any stronger than a power law in $\tom_{1}$
\bel{f6a}
h_{m}\left(\tom_{1},\tau\right) =  1 + \tilde{c}_{m}\tom_{1}^{\lambda_{m}(\tau) - 1}
\ee
where the monotonically increasing\footnote{\sf In Gibbon \etal (2014), in which only the maxima in time of $D_{m}$ were considered, 
there were indications that the $\lambda_{m}$ decreased with $m$, albeit very weakly, which is not consistent with the constraint 
$\lambda_{m} \leq \lambda_{m+1}$ that stems from (\ref{f4}). This suggests that the choice of $h_{m}$ made in (\ref{f6a})  needs 
a slight modification. This issue needs testing with a proper set of numerical experiments.} $\lambda_{m}$ must lie in the range
\bel{f5f}
1 \leq \lambda_{m}(\tau) \leq 4\,.
\ee
Morover,  the dimensionless constants $\tilde{c}_{m}$ will be chosen more specifically later. The final result is that if there 
exists a strong solution on the interval $[0,\,T^{*})$, then with the choice of $h_{m}$ as in (\ref{f6a}), there exists a set of exponents 
$\{\lambda_{m}(\tau)\}$ such that 
\bel{f6c}
\tom_{m} = \tom_{1}\left[1+ \tilde{c}_{m}\tom_{1}^{\lambda_{m}(\tau) - 1}\right]^{\frac{m-1}{m}}\,,\qquad 1 < m < \infty\,.
\ee
In practice $\tom_{1}$ is so large that we may safely ignore the factor of unity in $h_{m}(\tom_{1},\tau)$ in (\ref{f6c}). With 
the definition 
\begin{align}\label{f6b}
\begin{split}
 A_{m,\lambda_{m}(\tau)} &= \shalf\alpha_{m}\left[ 1 + (\lambda_{m}(\tau) - 1)\left(\frac{m-1}{m}\right)\right] = 
\frac{(m- 1)\lambda_{m}(\tau) + 1}{4m-3}\,,
\end{split}
\end{align}
we have
\bel{f5e}
D_{m} = c_{m} D_{1}^{A_{m,\lambda_{m}(\tau)}}\,.
\ee
which is a version of (\ref{f7}) with $\lambda_{m} = \lambda_{m}(\tau)$ as a function of time.  The introduction of the set 
$\{\lambda_{m}(\tau)\}$ means the dynamics are analyzed in terms of $D_{1}$ and $\{\lambda_{m}(\tau)\}$ instead of $D_{1}$ 
and $\{D_{m}(\tau)\}$. In (\ref{f7}) the inequality is appropriate because in Gibbon \etal (2014) $\lambda_{m}$ was estimated 
as a constant parameter corresponding to maxima in time in plots of $\ln D_{m}/\ln D_{1}$ versus time. Note that when 
$\lambda_{m} = 4$, $A_{m,4} = 1$, at which point the $D_{m}$ versus $D_{1}$ relation is linear. Beyond the range in (\ref{f5f}), 
when $\lambda_{m} > 4$, (\ref{f5e}) is no longer valid.  Clearly, a further set of numerical experiments are needed to analyze 
more closely the trajectories of $\lambda_{m}(\tau)$ in (\ref{f5e}). 
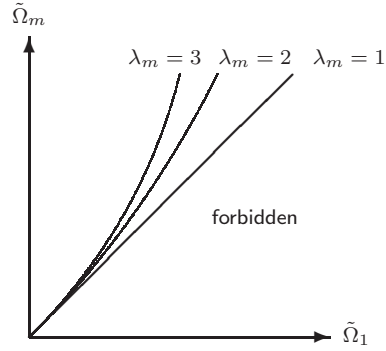
\begin{figure}
\centering
\setlength{\unitlength}{5mm}
\begin{picture}(10,10)
\thicklines
\put(0,0){\vector(0,1){8}}
\put(0,0){\vector(1,0){8}}
\thicklines
\put(0,0){\line(1,1){7}}
\put(0,8.3){\makebox(0,0)[b]{\scriptsize$\tom_{m}$}}
\put(8.7,-0.2){\makebox(0,0)[b]{\scriptsize$\tom_{1}$}}
\put(8.5,7.2){\makebox(0,0)[b]{\scriptsize$\lambda_{m} = 1$}}
\thinlines
\qbezier(0,0)(3,3)(5,7)
\put(6,7.2){\makebox(0,0)[b]{\scriptsize$\lambda_{m} = 2$}}
\qbezier(0,0)(3,3)(4,7)
\put(3.6,7.2){\makebox(0,0)[b]{\scriptsize$\lambda_{m} = 3$}}
\put(6,3){\makebox(0,0)[b]{\scriptsize\sf forbidden}}
\end{picture}
\caption{\sf\scriptsize Plots of the convex curves $\tom_{m}$ versus $\tom_{1}$ for $\lambda_{m} = 2,~3$ lying above the straight line 
$\lambda_{m} = 1$. As $\lambda_{m}$ increases the $\tom_{m}$ spread more, corresponding to greater intermittency in the 
$\bom$-field.}\label{Omplot}
\end{figure}

\section{\textsf{\color{blue}Navier-Stokes results in the $D_{m}-D_{1}$ plane parametrized by $\lambda_{m}(\tau)$}}

The essential idea behind (\ref{f6b}) has been to avoid the use of derivatives of $\bom$ and instead transfer the dynamical 
relationship behind $\tom_{m}$ and $\tom_{1}$ to the $\{\lambda_{m}(\tau)\}$. This allows us to treat the $D_{m}-D_{1}$ 
phase as a phase plane parametrized by $\lambda_{m}(\tau)$, as in Fig. 2. The curves of constant $\lambda_{m}$ are drawn 
as concave curves, although actual orbits $\lambda_{m}(\tau)$ could potentially wander across the phase plane and over 
these labelled boundaries, even though numerical experiments 
so far have found that their maxima lie in the lower half of the lowest tongue labelled by the dashed curves. We wish to 
demonstrate certain results about the nature of solutions in different sectors of this phase plane bounded by curves 
$\lambda_{m} = \mbox{const}$, so in this section $\lambda_{m}$ is treated as a constant. 
\par\smallskip
The curve $\lambda_{m} =1$ is associated with the lower bound  $\tom_{1} \leq 
\tom_{m}$, which translates to 
\bel{g1}
D_{1}^{\alpha_{m}/2} \leq D_{m}\qquad\mbox{where}\qquad A_{m,1} = \alpha_{m}/2 = \frac{m}{4m-3}\,.
\ee
Thus the regions below the line in Fig. 1 and below the lowest curve in Fig. 2, both corresponding to $\lambda_{m} =1$, are 
forbidden by H\"older's inequality. Moreover, the relation between  $D_{m}$ and $D_{1}$ is linear when $\lambda_{m} 
=4$. The results of \S\ref{lam12} alone can be found in Gibbon \etal (2014)\,: the rest of the material is new. 
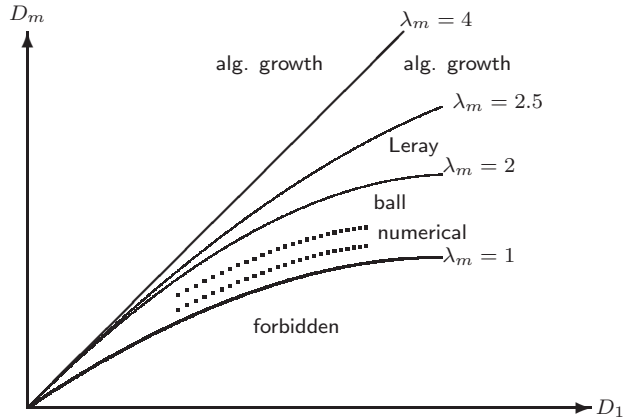
\begin{figure}
\centering
\setlength{\unitlength}{5mm}
\begin{picture}(11,11)
\thicklines
\put(0,0){\vector(0,1){10}}
\put(0,0){\vector(1,0){15}}
\thicklines
\put(0,0){\line(1,1){10}}
\put(10.86,10.1){\makebox(0,0)[b]{\scriptsize $\lambda_{m}  = 4$}}
\thinlines
\put(0,10.25){\makebox(0,0)[b]{\scriptsize$D_{m}$}}
\thinlines
\put(15.5,-0.25){\makebox(0,0)[b]{\scriptsize$D_{1}$}}
\put(12.5,7.9){\makebox(0,0)[b]{\scriptsize $\lambda_{m}  = 2.5$}}
\thinlines
\qbezier(0,0)(6,6)(11,8)
\thinlines
\qbezier(0,0)(6,6)(11,6.2)
\put(10,9){\makebox(0,0)[b]\textbf{\scriptsize\sf alg. growth}}
\put(5,9){\makebox(0,0)[b]\textbf{\scriptsize\sf alg. growth}}
\put(9.6,6.8){\makebox(0,0)[b]\textbf{\scriptsize\sf Leray}}
\put(9.2,5.3){\makebox(0,0)[b]\textbf{\scriptsize\sf ball}}
\put(6,2){\makebox(0,0)[b]\textbf{\scriptsize\sf forbidden}}
\put(12,6.2){\makebox(0,0)[b]{\scriptsize $\lambda_{m}  = 2$}}
\thicklines
\qbezier(0,0)(6,4)(11,4)
\put(12,3.8){\makebox(0,0)[b]{\scriptsize $\lambda_{m}  = 1$}}
\linethickness{.4mm}
\qbezier[25](4,3)(7,4.6)(9,4.8)
\qbezier[25](4,2.6)(7,4.1)(9,4.3)
\put(10.5,4.5){\makebox(0,0)[b]{\scriptsize\sf numerical}}
\end{picture}
\caption{\sf\scriptsize A cartoon in the $D_{1}-D_{m}$ plane parametrized by $\lambda_{m}$\,: in the region $2.5 \leq \lambda_{m} 
\geq 4$ the $D_{m}$ undergo no more than algebraic growth with a restriction on (large) initial data for $2.5 \leq \lambda_{m} \leq 4$\,;  
in the region $2 < \lambda_{m} < 2.5$ only Leray's weak solutions are known to exist with the only control over $D_{m}$ being the 
bounded long time averages $\left<D_{m}\right>_{T} < \infty$\,; at $\lambda_{m} =2$ the bound on $D_{1}$ grows exponentially in time\,;  
in the region $1 \leq \lambda_{m} < 2$ the $D_{m}$ lie within an absorbing ball. The dotted curves represent the approximate regions 
of the maxima in the numerical simulations reported in Donzis \etal (2013) and Gibbon \etal (2014). The region below the curve 
$\lambda_{m} =1$ is forbidden as H\"older's inequality is violated there.}\label{phase}
\end{figure}

\subsection{\textsf{\color{blue}The region $1 \leq \lambda_{m} < 2$ and the curve $\lambda_{m} =2$}}\label{lam12}

From the definition of the $D_{m}$ in (\ref{Dmdef1}) note that $D_{1} = L \nu^{-2}\|\bom\|_{2}^{2}$.  
A purely formal differential inequality for $D_{1}$ is
\bel{b2}
\shalf 
\dot{D}_{1} \leq L\nu^{-2}\left\{-\nu \I |\nabla\bom|^{2}\,dV + \I |\nabla\bu||\bom|^{2}\,dV + 
L^{-1}\|\bom\|_{2}\|\bdf\|_{2}\right\}\,.
\ee 
Dealing with the negative term first, an integration by parts gives
\bel{b3}
\I |\bom|^{2}\,dV \leq \left(\I|\nabla\bom|^{2}dV\right)^{1/2}\left(\I|\bu|^{2}dV\right)^{1/2}\,,
\ee
where the dimensionless energy $E$ is defined as
\bel{b4}
E = \nu^{-2}L^{-1}\I|\bu|^{2}\,dV\,.
\ee
This is always bounded such that
\bel{b7}
\overline{\lim}_{t\to\infty}E \leq c\,Gr^{2}\,.
\ee
Next, the nonlinear term in (\ref{b2}) needs to be estimated. The standard result using a Sobolev inequality produces a 
cubic nonlinearity $D_{1}^3$ that is too strong for the negative term\,: all that can be deduced from this is that $D_{1}$ 
is bounded from above only for short times or for small initial data. The difficulty caused by this term has been known 
for many decades\,: see Constantin and Foias (1988) and Foias \etal (2001). We circumvent this problem by proceeding 
as follows\,:
\begin{align}\label{b8a}
\begin{split}
\I |\nabla\bu||\bom|^{2}dV &=\I |\bom|^{\frac{2m-3}{m-1}}|\bom|^{\frac{1}{m-1}}|\nabla\bu|dV\\
&\leq \left(\I|\bom|^{2}dV\right)^{\frac{2m-3}{2(m-1)}}
\left(\I|\bom|^{2m}dV\right)^{\frac{1}{2m(m-1)}}\left(\I|\nabla\bu|^{2m}dV\right)^{\frac{1}{2m}}\\
&\leq  C_{m}\left(\I|\bom|^{2}dV\right)^{\frac{2m-3}{2(m-1)}}\left(\I|\bom|^{2m}dV\right)^{\frac{1}{2(m-1)}}\\
&= C_{m}L^{3}\varpi_{0}^{3} D_{1}^{\frac{2m-3}{2m-2}}D_{m}^{\frac{4m-3}{2m-2}}\,,\qquad\qquad 1 < m < \infty\,.
\end{split}
\end{align}
The penultimate line is based on $\|\nabla\bu\|_{p} \leq c_{p} \|\bom\|_{p}$, for $1 < p < \infty$. Inserting the depletion 
$D_{m} = c_{1,m}D_{1}^{A_{m,\lambda_{m}}}$ gives 
\bel{b8b}
L\nu^{-2}\I |\nabla\bu||\bom|^{2}\,dV \leq c_{2,m}\varpi_{0}\,D_{1}^{\xi_{m,}}\,,
\ee
where $\xi_{m}$ is defined as 
\bel{ximdefA}
\xi_{m} = \frac{A_{m,\lambda_{m}}(4m-3) +2m - 3}{2(m-1)}.
\ee
It can now be seen that by using (\ref{f7}), the $m$-dependency cancels leaving\footnote{\sf Lu and Doering (2008) 
showed numerically 
that by maximizing the enstrophy subject to $\mbox{div}\,\bu = 0$, two branches of the nonlinear term appear,  the lower 
being $D_{1}^{1.78}$ and the upper $D_{1}^{2.997}$. Later, Schumacher, Eckhardt and Doering (2010), suggested that 
$7/4$ and $3$ were the likely values of these two exponents\,: the exponent $1+ \shalf\lambda_{m} = 7/4$ corresponds to 
$\lambda_{m}  = 1.5$ which lies at the upper end of the range $1.15 \leq \lambda_{m} \leq 1.5$ observed in Gibbon \etal (2014).}
\bel{ximdeldef}
\xi_{m} = 1 + \shalf \lambda_{m}\,.
\ee
$\xi_{m}$ does not reach its conventional value of 3 unless $\lambda_{m} =4$. The differential inequality (\ref{b2}) now becomes 
($\tau = \varpi_{0}t$)
\bel{b9}
\shalf \frac{d D_{1}}{d\tau} \leq -\frac{D_{1}^{2}}{E} + c_{2,m} D_{1}^{1+ \shalf\lambda_{m}} + Gr D_{1}^{1/2}\,.
\ee
Given that $E$ is bounded above, $D_{1}$ is always under control provided $\lambda_{m}$ is restricted to the range 
$1 \leq \lambda_{m}< 2$. Formally, there exists an absorbing ball for $D_{1}$ of radius
\bel{D1ballB}
\overline{\lim}_{t\to\infty} D_{1}\leq \tilde{c}_{2,m}Gr^{\frac{4}{2 - \lambda_{m}}} + O\left(Gr^{4/3}\right)\,,
\ee
It has been shown in Gibbon \etal (2014) that this gives rise to a global attractor 
$\mathcal{A}$ whose Lyapunov dimension has been estimated as
\bel{ad5a}
d_{L}(\mathcal{A}) \leq \left\{
\begin{array}{l}
c_{4,m}Re^{\frac{3}{5}\left(\frac{6 - \lambda_{m}}{2-\lambda_{m}}\right)}\\
c_{5,m}Gr^{\frac{3}{5}\left(\frac{4-\lambda_{m}}{2-\lambda_{m}}\right)}
\end{array}\right.
\ee
depending on whether one chooses to use the Reynolds or Grashof  number.  In the limit $\lambda_{m}\to 2$  the radius of the 
ball in (\ref{D1ballB}) grows but, at $\lambda_{m} =2$, the finiteness of $\int_{0}^{t}D_{1}(\tau)d\tau$ means that $D_{1}(t)$ 
has an exponentially growing upper bound. 

\subsection{\textsf{\color{blue}The regions $2.5 \leq \lambda_{m} \leq 4$}}

It has also been shown in Gibbon (2012) and Gibbon \etal (2014) that the $D_{m}$ satisfy the differential inequality 
for $1 < m < \infty$
\bel{Dmdi1}
\dot{D}_{m} \leq D_{m}^{3}\left(-\varpi_{1,m}\left( \frac{D_{m}}{D_{1}}\right)^{\eta_{m}}
+ \varpi_{2,m}\right) + \varpi_{3,m}Gr D_{m}^{1-1/\alpha_{m}}\,,
\ee
where $\eta_{m} = 2m/3(m-1)$ and where $\varpi_{1,m} < \varpi_{2,m}$ are constant frequencies. The last (forcing) term, 
is hard to handle in conjunction with the others so this will be separated and dealt with last\,: no more than algebraic 
growth in time can come from it. Let us now divide (\ref{Dmdi1}) by $D_{m}^{3}$ to obtain 
\bel{Dmdi2}
\shalf \frac{d~}{dt} D_{m}^{-2} \geq \varpi_{1,m}X_{m}(t)D_{m}^{-2} - \varpi_{2,m}
\ee
where
\bel{Xmdef}
X_{m} = D_{m}^{2}\left(\frac{D_{m}}{D_{1}}\right)^{\eta_{m}}\,.
\ee
A bound away from zero of the time integral of $X_{m}(t)$ is required to show that $D_{m}^{-2}(t)$ never passes through 
zero for some range of initial conditions. To achieve this we introduce the nonlinear depletion as in (\ref{Dmdef1}). Noting that 
$\eta_{m} + 2 = 2(4m-3)/3(m-1) = 2\eta_{m}\alpha_{m}^{-1}$, it is found that ($\tilde{c}_{m} = c_{m}^{2+\eta_{m}}$)
\begin{align}\label{Xmint1}
\begin{split}
X_{m} &= \tilde{c}_{m}D_{1}^{\eta_{m}(\lambda_{m} -1)(m-1)/m}\\
&= \tilde{c}_{m}D_{1}^{2(\lambda_{m} -1)/3}\,.
\end{split}
\end{align}
It is at this point that we introduce the lower bound\footnote{\sf Doering and Foias (2002) have shown that there are 
two estimates for the lower bound to the integral in (\ref{DF02}). The first is proportional to $t\,Gr$ and the second 
to $t\,Gr^{2}Re^{-2}$ depending on the relative sizes of the forcing length scale $\ell$ and the Taylor micro-scale. 
The first has been derived using a Poincar\'e inequality so it is likely to be less sharp although the two coincide when 
the bound $Gr \leq c\,Re^{2}$ is saturated. For simplicity we use the $Gr$-bound.} on $\int_{0}^{t}D_{1}(t')\,dt'$ found 
in Doering and Foias (2002)
\bel{DF02}
\int_{0}^{t}D_{1}(t')\,dt' \geq t\,Gr + O(t^{-1})\,.
\ee
This comes into play only if $2(\lambda_{m} -1)/3 \geq 1$, in which case $\lambda_{m} \geq 5/2$, where we can then use 
a Schwarz inequality to obtain 
\begin{align}\label{Xmint2}
\begin{split}
\int_{0}^{t}X_{m}(t')\,dt' &\geq  \tilde{c}_{m} \left(\int_{0}^{t}D_{1}\,
dt'\right)^{2(\lambda_{m} -1)/3}\,t^{1 - 2(\lambda_{m} -1)/3}\\
&\geq t\,\tilde{c}_{m} Gr^{2(\lambda_{m} -1)/3} + O\left(t^{1 - 4(\lambda_{m} -1)/3}\right)\,.
\end{split}
\end{align}
(\ref{Dmdi2}) can be solved to give
\begin{align}\label{Xmint3}
\begin{split}
\shalf [D_{m}(t)]^{2} &\leq \frac{\exp\{-\varpi_{1,m}\int_{0}^{t}X_{m}(t')\,dt'\}}{\shalf[D_{m}(0)]^{-2} 
- \varpi_{2,m}\int_{0}^{t}\exp\{- \varpi_{1,m}\int_{0}^{t'}X_{m}(t'')\,dt''\}dt'}\\
&= \frac{\exp\{-c_{m}\varpi_{1,m}tGr^{2(\lambda_{m} -1)/3}\}}
{\shalf[D_{m}(0)]^{-2} - \varpi_{2,m}\left[\tilde{c}_{m}\varpi_{1,m}Gr^{2(\lambda_{m} -1)/3}\right]^{-1}
\left(1 - \exp\{-\tilde{c}_{m}\varpi_{1,m}t\,Gr^{2(\lambda_{m} -1)/3}\}\right)}\,.
\end{split}
\end{align}
(\ref{Xmint3}) cannot develop a zero in the denominator if 
\bel{id1}
D_{m}(0) \leq \left(\shalf \tilde{c}_{m} \varpi_{1,m}\varpi_{2,m}^{-1}\right)^{1/2}
Gr^{(\lambda_{m} -1)/3}\,,
\ee
for any $5/2 \leq \lambda_{m} \leq 4$, in which case the solution decays exponentially.  This initial data is not small but it is not huge either. 
The estimates above have been achieved by neglecting the forcing term in (\ref{Dmdi1}). The effect of this term on its own yields
\bel{alg1B}
D_{m}(t) \leq \left[\alpha_{m}\varpi_{3,m}Gr \left( t_{0} + t \right)\right]^{\alpha_{m}}\,.
\ee

\subsection{\textsf{\color{blue}The region $2 < \lambda_{m} < 2.5$}}\label{lam225}

The dynamics in the middle tongue-shaped region bounded by the two central curves $\lambda_{m} = 2$ and $\lambda_{m} = 2.5$ 
in Fig. 2 remains open. Neither the depletion of nonlinearity nor the increase in dissipation afforded by (\ref{f5e}) are strong 
enough so we must fall back on the existence of Leray's weak solutions.The lower bound of Doering and Foias (2002) on 
the time integral of the enstrophy can only be used on the time integral (see (\ref{Xmint2})
\bel{D1int1}
\int_{0}^{t}D_{1}^{2(\lambda_{m} -1)/3}dt'
\ee
when $\lambda_{m} \geq 2.5$. For the range $2 < \lambda_{m} < 2.5$ the dimensionless energy $E$ can instead be used to bound 
(\ref{D1int1}) below, but if it passes close to zero for long enough, in the manner of a homoclinic orbit, then the resulting 
lower bound may be too small to be of use. However, (\ref{D1int1}) could be estimated numerically to monitor its behaviour.

\subsection{\textsf{\color{blue}The region $\lambda_{m} \geq 4$}}

The relation between $D_{m}$ and $D_{1}$ expressed in (\ref{f5e}) is valid only for $1 \leq \lambda_{m} \leq 4$. 
At $\lambda_{m} = 4$ the relation is linear; namely $D_{m} = c_{m}D_{1}$. For the region 
\bel{g4a}
D_{m} \geq c_{m}D_{1}
\ee
we can insert this relation directly into (\ref{Dmdi1}) and, provided the $c_{m}$ in (\ref{f5e}) are chosen such that $c_{m} \geq 
\varpi_{2,m}\varpi_{1,m}^{-1}$, (\ref{Dmdi1}) reduces to 
\bel{g4b}
\dot{D}_{m} \leq \varpi_{3,m}Gr D_{m}^{1-1/\alpha_{m}}\,,
\ee
yielding the same bound for $D_{m}(t)$ as in (\ref{alg1B}).

\section{\textsf{\color{blue}Conclusion}}

Subject to the closure relation (\ref{f5e}), in which the effect of the higher scaled norms $D_{m}$ is hidden in the evolution of 
the set of functions $\lambda_{m}(\tau)$, it has been shown that the $3D$ Navier-Stokes equations are regular in each of the 
regions when $\lambda_{m} \geq 1$, with the exception of the region $2 < \lambda_{m} < 2.5$. There is also the proviso that 
initial data is limited in the range $2.5 < \lambda_{m} \leq 4$. For initial data set in the region $\lambda_{m} \geq 4$ (above 
the straightline in Fig 2), the 
dynamics are no worse than algebraic growth, although such initial data is highly pathological. These results are summarized 
in Fig. 2. The three regular regions are fundamentally different. Solutions lying in the region $\lambda_{m} \geq 2.5$ seem 
moribund in the sense that the forcing dominates 
only algebraically. However, solutions in the lower region or tongue $1\leq \lambda_{m} < 2$ live in an absorbing ball, 
the radius of which is given in (\ref{D1ballB}), and it is here where all the interesting dynamics lies. It has been shown in 
Gibbon \textit{et~al} (2014) that solutions here have a corresponding spectrum that is consistent with statistical turbulence 
theories\,: see Frisch (1995) and Doering and Gibbon (2002). As drawn in Fig. 2 (dotted 
curves), the large-scale numerical simulations reported in Gibbon \etal (2014) lie well within this region. This is only 
partially satisfactory  (see below) in the sense that the existence of an absorbing ball is enough for the existence of a 
global attractor \textit{provided the solution trajectory $\lambda_{m}(\tau)$ remains in this region}. There are two 
alternatives\,: 
\ben
\item Orbits that originate in the range $1 \leq \lambda_{m} \leq 2$ always remain there\,;
\item Orbits originating in  the range $1 \leq \lambda_{m} \leq 2$ could travel out of this region and into the range 
$2 < \lambda_{m} < 2.5$ and beyond. However, the nature of this transition is uncertain, and it is unclear what the nature 
of weak solutions would mean numerically if this happened. 
\een
The numerical simulations performed so far have all had their initial data resting in $1 \leq \lambda_{m} \leq 2$ and have 
shown no evidence for the behaviour in item 2. In fact, the observed range $1.15 \leq \lambda_{m} \leq 1.5$ indicates 
relatively mild dynamics. Unless a rigorous proof is found for the behaviour in item 1, the possibility that the behaviour in 
item 2 could occur for higher values of $Re$ ought to be kept in mind. A series of numerical experiments are needed with 
initial conditions set in the four different regions which track the evolution of $\lambda_{m}(\tau)$ although in the scaled 
time $\tau = \varpi_{0}t$, $\varpi_{0}$ could be so small that one may have to compute for significantly large values of the 
real time $t$. If the behaviour in item 2 is observed this would open the question of the physical manifestation of weak 
solutions. Given that these solutions lack uniqueness, would there be there a corresponding physical effect, such as multiple 
branching of the $\lambda_{m}$-trajectories? 

\section*{\textsf{\color{blue}Acknowledgment}}
Thanks are due to Darryl Holm for discussions on the nature of the $\tom_{m}$. 

\appendix
\section{\textsf{\color{blue}The triangular H\"older inequality (\ref{Hin2})}}\label{app}

Consider the definition of $\Omega_{m}$ 
\bel{app1}
L^{3}\Omega_{m}^{2m} = \I |\bom|^{2m}\,dV \equiv \I|\bom|^{2\alpha}|\bom|^{2\beta}dV
\ee
where $\alpha + \beta = m$. Then, for $m > 1$ and $1 \leq p \leq m-1$ and $q > 0$, we have 
\bel{app2}
L^{3}\Omega_{m}^{2m} \leq  \left(\I |\bom|^{2(m-p)}\,dV\right)^{\frac{\alpha}{m-p}}
\left(\I|\bom|^{2(m+q)}dV\right)^{\frac{\beta}{m+q}}
\ee
where $\frac{\alpha}{m-p} + \frac{\beta}{m+q} = 1$. Solving for $\alpha,~\beta$ gives 
\bel{app3}
\alpha = \frac{q(m-p)}{p+q}\qquad\mbox{and}\qquad \beta = \frac{p(m+q)}{p+q}\,,
\ee
thereby giving
\bel{app4}
\Omega_{m}^{m(p+q)} \leq \Omega_{m-p}^{q(m-p)}\Omega_{m+q}^{p(m+q)}\,.
\ee
Now choose $q=1$ and $p = m-1$ to obtain
\bel{app5}
\Omega_{m}^{m^2} \leq \Omega_{1}\Omega_{m+1}^{m^{2}-1}\,,
\ee 
which leads to (\ref{Hin2}). 


\end{document}